\documentclass[12pt]{article}

\usepackage{epsf}
\usepackage{amssymb}

\usepackage{graphicx}
\usepackage[figuresright]{rotating}

\newcommand{\rf}[1]{(\ref{#1})}
\newcommand{\beq}{\begin{equation}}
\newcommand{\eeq}{\end{equation}}
\newcommand{\bea}{\begin{eqnarray}}
\newcommand{\eea}{\end{eqnarray}}

\newcommand{\e}{\mbox{e}}
\renewcommand{\d}{\mbox{d}}
\newcommand{\g}{\gamma}

\renewcommand{\l}{\lambda}
\renewcommand{\L}{\Lambda}
\renewcommand{\b}{\beta}
\renewcommand{\a}{\alpha}
\newcommand{\n}{\nu}
\newcommand{\m}{\mu}

\renewcommand{\th}{\theta}
\newcommand{\ep}{\varepsilon}

\newcommand{\del}{\delta}

\newcommand{\sg}{\sigma}
\renewcommand{\k}{\kappa}

\newcommand{\ra}{\rangle}
\newcommand{\la}{\langle}
\newcommand{\prt}{\partial}
\newcommand{\mi}{\!-\!}
\newcommand{\equ}{\!=\!}
\newcommand{\pl}{\!+\!}

\newcommand{\cD}{{\cal D}}

\newcommand{\cN}{{\cal N}}

\begin{document}

\begin{center}
\vspace{24pt}
{ \large \bf 
Simplicial Euclidean and Lorentzian Quantum Gravity}\footnote{Plenary 
talk at GR16, Durban, July 2001.}

\vspace{30pt}

{\sl J. Ambj\o rn}

\vspace{24pt}
{\footnotesize

The Niels Bohr Institute, \\
Blegdamsvej 17, DK-2100 Copenhagen \O , Denmark\\
{\it email: ambjorn@nbi.dk}

}
\vspace{48pt}

\end{center}


\begin{center}
{\bf Abstract}
\end{center}

One can try to define the theory of quantum gravity as the 
sum over geometries. In two dimensions the sum over 
{\it Euclidean} geometries can be performed constructively
by the method of {\it dynamical triangulations}. 
One can define a  {\it proper-time} propagator. This 
propagator can be used to calculate generalized
Hartle-Hawking amplitudes and it can be used 
to understand the 
the fractal structure of {\it quantum geometry}.
In higher dimensions the philosophy of defining the 
quantum theory, starting from a sum over Euclidean geometries, 
regularized by a reparametrization invariant cut off 
which is taken to zero, seems not to lead to an interesting
continuum theory. The reason for this is 
the dominance of singular Euclidean
geometries. Lorentzian geometries 
with a global causal structure are less singular.
Using the framework of dynamical triangulations
it is possible to give a constructive definition 
of the sum over such geometries, In two dimensions
the theory can be solved analytically. It
differs from two-dimensional Euclidean quantum gravity,
and the relation between the two theories can be 
understood. In three dimensions the theory avoids 
the pathologies of three-dimensional Euclidean quantum 
gravity. General properties of the four-dimensional 
discretized theory have been established, but a
detailed study of the continuum limit in the spirit
of the renormalization group and {\it asymptotic safety}
is till awaiting.

\vspace{12pt}
\noindent


\newpage

\section{Introduction}

We are still searching for the theory of quantum 
gravity. While the four dimensional theory of quantum 
gravity may have to go beyond the conventional 
framework of quantum field theory it is possible 
to discuss two- and three-dimensional quantum 
gravity in terms of conventional field theoretical 
concepts. In this way one may hope to learn 
important lessons about the ``real'' theory 
of four-dimensional gravity, whether or not it can be defined 
as a non-perturbative quantum field theory.

Two-dimensional quantum gravity is an example
of the subtlety present in a reparameterization 
invariant theory. In two dimensions 
the Einstein action is trivial as long as we do not sum over 
different topologies of {\it space-time} (which we are 
not attempting here). Thus we are left only with the 
cosmological term in the action and the classical 
theory is trivial. Diffeomorphism invariance ensures
that no {\it field} degrees of freedom exist.
Nevertheless, a finite number of quantum mechanical
degrees of freedom survives, which still
have a rich (quantum) geometrical description.
Despite the fact that no field degrees of freedom 
exist, it still makes sense to talk about generalized 
Hartle-Hawking like amplitudes: the sum over all 
Euclidean two-geometries with boundaries of
of lengths $L_k$, $k \equ 1,\ldots,n$. Similarly 
one can define the concept of correlators, depending on 
geodesic distance and show that standard concepts, derived 
from Euclidean quantum field theory or the theory 
of critical phenomena, make sense even in a framework
of fluctuating geometries. 

Three-dimensional quantum gravity can be addressed
in the same spirit. The new aspects of three-dimensional 
quantum gravity compared to two-dimensional quantum gravity 
are the following:  the theory is a perturbative non-renormalizable 
theory in the gravitational coupling constant and the  Euclidean 
action is unbounded from below. It is thus {\it a priori} unclear 
how to make sense of an Euclidean three-dimensional path 
integral. On the other hand we know that the conformal factor 
only appears as a constraint in a reduced phase-space 
quantization and thus should not really pose a problem
in a Lorentzian quantization of three-dimensional quantum
gravity. It is of interest to understand if it is possible
at all to think about three-dimensional quantum gravity 
as a path integral involving
the sum over a certain class of three-dimensional Euclidean geometries of 
a given topology. 

Contrary to the situation in two dimensions 
this ``purely'' Euclidean approach has not been 
successful in three dimensions. This led to the 
concept of ``Lorentzian'' gravity, the idea being that 
only Lorentzian geometries with a global causal structure
should be included in the path integral, and that a possible
rotation to Euclidean geometries should respect this.
Returning to two-dimensional quantum gravity where analytic 
solutions can be found, one can implement this program 
and one finds that  Lorentzian quantum gravity 
{\it is} different from Euclidean quantum gravity. Using 
proper-time as an evolution parameter, two-dimension Euclidean 
quantum gravity can be understood as two-dimensional 
Lorentzian quantum gravity where it is possible to create 
at each space-time point  a baby universe, 
which at later proper-time 
remains separated from the ``parent'' universe. In three-dimensions 
the path-integral over Lorentzian geometries can be studied by 
numerical as well as analytical methods, and it seems an 
interesting candidate for a theory of quantum gravity, defined 
as the sum over geometries, with relative weights dictated by the 
Einstein-Hilbert action.

\section{Non-perturbative regularization}\label{nonpert}

In a theory of quantum gravity it is not enough to 
study perturbation theory. The examples of the $\phi^4$ field 
theory and $QCD$ in four dimensions illustrates this.
Both theories have well defined perturbative expansions,
and observables can be calculated to any order in the 
coupling constant in both theories. Since both theories are 
renormalizable one can express the observables entirely 
in terms of the renormalized mass and coupling constant.
However, it does not ensure that a genuine quantum field 
exists without a cut-off. In order to address this question 
one first has to define the theory outside perturbation theory
and then show that observables calculated in 
the non-perturbatively defined theory become independent of 
a possible cut-off, used in the definition of the non-perturbative
framework. Finally, one can then discuss to what extend the perturbation
expansion, which is usually at most an asymptotic expansion, contains
information about the non-perturbatively defined theory.

All the evidence suggests that $QCD$ exists as a genuine quantum 
field theory in four-dimensional space-time. One needs 
a cut-off at an intermediate step when defining the theory,
but this cut off will never show up in any continuum physical 
observables. The situation is opposite for the $\phi^4$ theory.
It seems impossible to define a non-trivial theory where 
the renormalized coupling constant is different from zero 
when the cut-off is removed. This situation is of course 
already hinted by low order perturbation theory where one observes 
a Landau pole at large energies. 

In the case of a $\phi^4$ theory, as it appears for instance in the 
Standard Model, one considers it as an effective low energy approximation
to a more more elaborate, yet to be understood, theory at the GUT scale 
or the Planck scale. Clearly, if we want to apply conventional 
quantum field theoretical concepts to the theory of quantum gravity,
such a situation is not good enough: it has to be a quantum 
theory independent of cut-off in the same way as $QCD$, since 
the quantum phenomena we want to consider take place 
precisely at the Planck scale. In order to understand whether or 
not the theory exists as a conventional quantum theory a non-perturbative
framework is thus mandatory.

For non-Abelian gauge theories and the $\phi^4$ scalar field theory
the use of a space-time {\it lattice} has provided a useful regularization,
in particularly when combined with the use of Monte Carlo simulations,
taking advantage of the fast present days computers. It has allowed us 
to related Euclidean quantum field theory to {\it the statistical 
theory of critical phenomena}. Stated very shortly Euclidean quantum 
field theories can be extracted from second order phase transitions 
of generalized spin systems defined on the lattice. The way the 
continuum field theory is recovered from the lattice spin model is 
as follows: by fine-tuning the coupling constants of the model , generically 
denoted $\l$, one can approach the phase transition point $\l_c$. The spin-spin
correlation length $\xi$, measured in lattice spacings, will diverge as 
\beq\label{2.1}
\xi(\l) \sim \frac{1}{|\l -\l_c|^{\n}}.
\eeq 
The lattice spacing $a$ serves as the cut-off. By Fourier transformation
it is seen that the (lattice) momentum must be less that $\pi/a$.
We introduce the physical  correlation length $L$ as $L \equ \xi(\l)\cdot a$.
One now take the limit 
\beq\label{2.2}
L = \xi(\l)\cdot a ~~~{\rm fixed},~~~\xi(\l)\to \infty,~~~~a \to 0.
\eeq
From eq.\ \rf{2.1} we see that the interpretation of this 
continuum limit is that while the correlation length measured 
in lattice units diverges when $\l \to \l_c$, the lattice 
spacing (the cut-off) goes to zero as:
\beq\label{2.3}
a (\l) \sim |\l-\l_c|^\n~~~{\rm for}~~~\l \to \l_c.
\eeq
In this {\it scaling limit} the lattice becomes increasingly
fine-grained compared to the physical correlation length $L$.
The lattice structure becomes unimportant for the physics 
associated with these long range fluctuations and Euclidean 
invariance will be restored for the physics related to these 
fluctuations. Let $\phi(x_n)$ denote the spin at lattice point
$x_n \equ n\cdot a$. 
Since the spin correlation functions will behaves
\beq\label{2.4}
\la \phi(x_n) \phi(0) \ra \sim \e^{-n/\xi(\l)} \sim \e^{-x_n/L}
\eeq
for large $n$, we see that the {\it renormalized physical mass}
should be identified with $1/L$:
\beq\label{2.5}
m_{ph} = 1/L.
\eeq
The choice of physical correlation length in \rf{2.2} is 
equivalent to a choice of renormalized mass in the 
continuum quantum field theory. We have used $\l$ as a 
generic coupling constant and usually the ``spin'' system 
will have a number of coupling constants. Combined fine-tuning 
will allow to fix in addition the other renormalized coupling 
constants of the continuum theory. Finally the 
lattice ``spin'' $\phi(x_n)$ is assigned a dimension from the 
short distance behavior of the spin-spin correlation 
function. In eq.\ \rf{2.4} it was assumed that 
$n \gg \xi(\l)$. In the opposite limit $1 \ll n \ll \xi(\l)$ one expects
\beq\label{2.6}
\la \phi(x_n) \phi(0) \ra \sim n^{-d+2+\eta} \sim 
\frac{a^{d-2+\eta}}{x_n^{d-2+\eta}},~~~~a \ll x_n \ll L.
\eeq
The factor $a^{d-2+\eta}$ reflects the scaling dimension of 
the {\it physical} field 
\beq\label{2.7}
\phi_{phys} = a^{-(d-2+\eta)/2} \phi_{lattice},
\eeq
and the {\it anomalous scaling dimension} $\eta$ is related
to wave function renormalization.

This way of defining a quantum field theory is closely 
related to use of the renormalization group in quantum field 
theory and it emphasizes the concept of {\it universality}:
many different ``spin'' systems defined on lattices will 
lead to the same class of continuum field theories, 
characterized by {\it critical exponents} like $\n$ and $\eta$.
One can imagine an idealized spin system where all possible 
spin interactions are included in the spin Hamiltonian. The 
key assumption is then that in this infinite dimensional 
coupling constant space the critical surfaces where one 
can define  continuum theories are of {\it finite co-dimensions}.
Thus there is only a finite number of coupling constants which 
need to be fine-tuned in order that the system becomes critical.
Translated to continuum physics only a finite number 
of coupling constants need renormalizations. In the 
context of a perturbative expansion around free field theory
this leads to the usual concept of renormalizable theories.
But the philosophy also applies to expansions about 
non-trivial critical points where there exist no 
free fields. Since four-dimensional quantum 
gravity is not renormalizable viewed as a 
perturbation theory around flat space, the simplest possible 
scenario still using the framework of conventional 
quantum field theory is that of a non-trivial fixed point
in the sense described above. In the context of quantum gravity 
this was first emphasized 
by Weinberg, who called it ``asymptotic safety'' since 
only a finite number of coupling constants needed 
adjustment in order to recover continuum physics.

While the lattice formulation of quantum field theories 
has the virtues mentioned above it also has a number 
of drawbacks:
\begin{itemize}
\item[(1)] The lattice formulation is usually not a convenient 
framework for analytic calculations.
\item[(2)] space-time symmetries are explicitly broken. In 
particular, the lattice seems not to be the best 
regularization for theories with local space-time symmetries.
\end{itemize}
{\it It is no necessarily so.} Two-dimensional Euclidean 
quantum gravity provides a counter example. There exists a 
simple lattice regularization of 2d Euclidean quantum 
gravity, called {\it dynamical triangulation}, which is 
\begin{itemize}
\item[(1)] convenient for analytic calculations,
\item[(2)] has no problems recovering the diffeomorphism 
invariant continuum limit,
\item[(3)] is defined directly on the space of geometries,
\item[(4)] serves as a  textbook example of universality when 
viewed as a statistical field theory.
\end{itemize}  

The use of dynamical triangulations \footnote{The concept ``Dynamical 
triangulations'' was  introduced in \cite{kkm,david,adf,adfo} mainly   
in an attempt to provide a non-perturbative definition of the bosonic string.}
is an attempt to 
approximate {\it the space of geometries} by the class of 
piecewise linear geometries which can be constructed from 
{\it an elementary building block}. In two dimensions the 
building block is an equilateral triangle and the piecewise 
linear geometries are obtained by gluing together 
the triangles in all possible ways consistent with a 
given two-dimensional topology. If we consider the triangles as 
flat, the (piecewise linear) geometry is uniquely determined 
by the connectivity pattern of the triangulation. The length 
$a$ of the links serves as the cut-off as for a regular lattice,
and with the interpretation given here it is a diffeomorphism
invariant cut-off, since we work directly with geometries.
The hope is that when $a \to 0$ this set of geometries will 
be ``dense'' in the set of all geometries which enters into 
the path integral. Also, it is important to understand that 
in the spirit of universality there is nothing special about the 
use of equilateral triangles. For instance, one could have used 
squares as building blocks, and if we wanted a piecewise
linear geometry  associated with such generalized ``quadrangulations''
one could subdivide the square into two triangles at one of the 
diagonals, considering each triangle as ``flat''. The results we will 
mention in the following will indeed be independent of such details.

Using a lattice to define the path integral also includes 
defining the discretized action. In conventional 
lattice theories the action is usually chosen such that 
it for $a \to 0$ converges to the classical continuum action of 
the field theory one tries to quantize. There is a large 
freedom in choosing such discretized actions, all leading to the
same continuum theory by the universality mentioned above. 
In the case of piecewise linear geometries the Einstein-Hilbert
action can actually be implemented entirely in terms 
of the piecewise linear geometry, as first noted by Regge \cite{regge}.
For a piecewise linear d-dimensional geometry the integrated
curvature is 
\beq\label{2.8}
\int \d^d \xi \sqrt{g}\, R = \sum_{\sg_{d-2}} V(\sg_{d-2})\, \ep (\sg_{d-2}), 
\eeq
where $\sg_{d-2}$ denotes a $(d\mi 2)$-dimensional simplex in the 
$d$-dimensional triangulation, $V(\sg_{d-2})$ the volume of the 
simplex and $\ep(\sg_{d-2})$ the deficit angle associated with the 
simplex. Eq.\ \rf{2.8} is ``exact'' in the sense that it {\it is} the 
natural curvature one can associate to a piecewise linear geometry.
As an example it leads to Eulers formula in the case of two-dimensional 
manifolds. 
 
In our case it becomes even simpler because the 
piecewise linear geometries are constructed from a 
fundamental building block. The space-time volume is thus 
proportional to the total number of building blocks and the 
curvature associated with a $(d\mi 2)$-simplex $\sg_{d\mi 2}$ is  related 
to the number $o(\sg_{d\mi 2})$ of 
$d$-simplices sharing it in the following way:
call the angle in a $d$-simplex between two $(d\mi 1)$-simplices 
sharing a $(d-2)$-simplex $\th$. Then the deficit angle associated with 
$\sg_{d\mi 2}$ is 
\beq\label{2.9}
\ep(\sg_{d\mi 2}) = 2\pi - \th \, o(\sg_{d\mi 1}).
\eeq
The Einstein-Hilbert action
\beq\label{2.9a}
S[g]= -\frac{1}{G} \int \d^d \xi \sqrt{g} R+ \L \int \d^d \xi \sqrt{g}
\eeq
associated with this kind of 
piecewise linear geometries is thus very simple and 
can be expressed as 
\beq\label{2.10}
S_T = -k_{d-2} N_{d-2}(T)+ k_d N_d,
\eeq  
where $N_{d-2}(T)$ denotes the number of $(d\mi 2)$-simplices and 
$N_d(T)$ denotes the number of $d$-simplices of the triangulation $T$.
The dimensionless coupling constants $k_{d\mi 2}$ and $k_d$ 
are related to the coupling constants $G$ and $\L$ as
\beq\label{2.11}
k_{d-2} \sim \frac{a^{d-2}}{G},~~~~~ k_d \sim \frac{a^{d-2}}{G}+a^d \L.
\eeq
This has the implication that in any dimension the calculation of 
the functional integral associated with the Einstein-Hilbert action 
is purely combinatorial:
\beq\label{2.12}
Z(G,\L) = \int \cD [g_{\m\n}] \; \e^{-S[g]} ~~\to~~ Z(k_{d\mi 2},k_d)=
\sum_T \e^{-S_T},
\eeq
where the sum can be rewritten as 
\beq\label{2.13}
Z(x,y) = \sum_{N_d,N_{d\mi 2}} x^{N_d} y^{N_{d\mi 2}} \cN(N_d,N_{d\mi 2}).
\eeq  
In this formula $x \equ e^{k_{d\mi 2}}$, $y\equ  e^{-k_d}$ and 
$\cN(N_d,N_{d\mi 2})$ denotes the {\it number} of different piecewise 
linear geometries one can construct from $N_d$ simplices, having $N_{d\mi 2}$
$(d\mi 2)$-simplices. In this way the partition function $Z(x,y)$ becomes
the generating function for the numbers  $\cN(N_d,N_{d\mi 2})$
as it is used in standard combinatorial analysis. This kind of 
remarkable simplification is only possible because we have 
restricted the sum over geometries to the ones with can be 
constructed from the cut-off size building blocks and because the 
Einstein action by the Regge-prescription is very simple for piecewise 
linear geometries.
 
\subsection{The continuum limit}

The combinatorial aspect of the sum over geometries
allows us to understand how to approach the continuum limit
in the simplest situations: the number of geometries of 
a given space-time topology and space-time volume grows
exponentially with space-time volume:
\beq\label{2.14}
\sum_{N_{d\mi 2}} \cN (N_{d\mi 2},N_d) \sim \e^{k_d^c N_d + o(N_d)}.
\eeq
From this it follows that {\it for each value of $k_{d\mi 2}$ there 
exists a critical value $k_d^c(k_{d\mi 2})$} such that 
\begin{itemize}
\item[(1)]
For $k_d > k_d^c(k_{d\mi 2})$ the partition function $Z(k_{d\mi 2},k_d)$
is convergent, while it is divergent for $k_d < k_d^c(k_{d\mi 2})$.
\item[(2)]
The {\it infinite volume limit} is obtained by fine-tuning $k_d \to 
k_d^c(k_{d\mi 2})$. This can be viewed as an additive renormalization
of the bare ``cosmological'' constant:
\beq\label{2.15}
k_d =   k_d^c(k_{d\mi 2}) + \L a^d,
\eeq
where $\L$ is viewed as a renormalized cosmological constant.
\item[(3)]
It might require an additional 
fine-tuning of $k_{d\mi 2}$ to obtain a {\it continuum limit}, 
if it exists at all. 
Such a fine-tuning can be viewed as a renormalization of the gravitational
coupling constant. (The analogy with the spin models might be helpful: 
the infinite volume limit means that the lattice is infinite, but it does 
not imply that we have a continuum field theory. That might require a 
fine-tuning of the coupling constants of the spin model). 
\end{itemize}
If a continuum limit of the kind discussed above can be 
constructed, the corresponding quantum field theory might serve as a
candidate for the {\it asymptotic safe} non-renormalizable theory 
in the sense of Weinberg.

\section{Two-dimensional Euclidean quantum gravity}\label{two-d}

\subsection{Hartle-Hawking wave functions}

Two-dimensional gravity, defined via dynamical triangulations, 
serves as a test of some of the above mentioned ideas. 
The integral of the curvature term is a topological invariant
in two dimensions. Thus, as long as we do not consider 
topology changes of space-time we can ignore the curvature term
in two dimensions and we are left with the cosmological term.
We write:
\beq\label{3.1}
Z_a(k_2) = \sum_T \e^{-k_2 N_2(T)} = \sum_{N} \e^{-k_2 N} \cN(N),
\eeq
where $\cN(N)$ denotes the number of different triangulations
constructed from $N$ equilateral triangles with link length $a$.
Objects of interest in the context of quantum gravity are the 
(generalized) Hartle-Hawking wave functions where the geometry 
of spatial boundaries are kept fixed, and the amplitude for 
such a configuration is obtained by summing over all two-dimensional 
geometries which have the prescribed spatial boundary geometries.
In two dimensions the geometry of the boundary is uniquely 
fixed by its length. In the regularization mentioned above a boundary
will consist of $l$ links of length $a$ from the adjacent triangles.
The length of the boundary is thus $L \equ l\cdot a$. The regularized 
Hartle-Hawking wave function can thus be written as
\beq\label{3.2}
G_a(l_1,\ldots,l_n;k_2) = \sum_{T(l_1,\ldots,l_n)} \e^{-k_2 N_2(T)}
= \sum_N \e^{-k_2 N} \cN_{l_1,\ldots,l_n}(N),
\eeq
where $\cN_{l_1,\ldots,l_n}(N)$ denotes the number of different 
 triangulations constructed from $N$ triangles and with 
$n$ boundaries consisting of  $l_1,\ldots,l_n$ links, respectively.

It is seen that the calculation of $Z_a(k_2)$ and $G_a(l_1,\ldots,l_n;k_2)$
can be reduced to purely combinatorial problems: the counting of 
distinct triangulations of various kinds. One finds, using either 
combinatorial methods or so-called matrix models, that 
\beq\label{3.3}
\cN (N) \sim \e^{k_2^c N} N^{\g -3},
\eeq
where $\sim$  means the leading asymptotic behavior when $N$ is large, and 
where $\g$ depends on the topology of two-dimensional space-time.
For the simplest topology of a sphere $\g\equ -1/2$. Thus we find
\beq\label{3.4}
Z_a(k_2) \sim \sum_N \e^{-(k_2-k_2^c)N}\; N^{\g-3} \sim  
\frac{1}{(k_2-k_2^c)^{\g-2}}.
\eeq
One observes a singular structure for $k_2 \to k_2^c$ which governs the 
large $N$ behavior of the sum, and a continuum limit can be 
obtained by an additive renormalization of the cosmological constant:
\beq\label{3.5}
k_2= k_2^c + a^2\L.
\eeq
An additive renormalization is expected since $\L$ has a positive 
mass dimension. The renormalization is ``natural'' in the sense that 
\beq\label{3.6}
(k_2-k_2^c)N = (\L a^2)N = \L (a^2 N) = \L \,V,
\eeq
where $V$ denotes the continuum space-time volume. The partition function
becomes
\beq\label{3.7}
Z_a(k_2) \sim (a^2 \L)^{2-\g} \sim a^{2-\g} \cdot Z(\L),~~~~Z(\L)= \L^{2-\g}.
\eeq
The divergent $a$-factor can be viewed as the kind of wave function 
renormalization which is always present in the path integral version 
of quantum field theory.

The counting can also  be performed in the case of triangulations with 
boundaries. The result is as  follows:
\beq\label{3.8}
\cN_{l_1,\ldots,l_n}(N) \sim \e^{k_2 N}\e^{k_b(l_1+\cdots + l_n)} 
N^{\g_n} (l_1\cdots l_n)^{\g_b} F(l_i/\sqrt{N}).
\eeq
The exponential growth with $N$ is counteracted by a renormalization
of the cosmological constant as above while the exponential growth with 
respect to the boundary length can be controlled by adding a boundary 
cosmological constant  to the action. This should anyway be done 
when we have space-like boundaries (see \cite{book} for details).
In this way we can take a continuum limit by scaling $l_i \sim \sqrt{N}$,
which is what one expects from the canonical dimensions of boundaries 
and bulk in two-dimensional quantum gravity. The resulting continuum
Hartle-Hawking wave functions can be found (see \cite{book,ajm}). We list the 
results in the case three boundaries (which is particularly simple):
\beq\label{3.9}
G(L_1,L_2,L_3,\L) = \L^{-1/2} \sqrt{L_1 L_2 L_3} \; 
\e^{-\sqrt{L}(L_1+L_2+L_3)}.
\eeq

\subsection{Universality}

It is important to understand that the results mentioned above are 
{\it universal}. The precise nature of the short distance 
regularization used is not important. We could have constructed 
our two-dimensional complexes from any combination of 
triangles, squares, pentagons etc with relative weights $g_3,g_4,g_5,\ldots$
and we would have obtained the same continuum Hartle-Hawking wave function
as long as the relative weights of the various kinds of polygons
are non-negative. If some of the weights are taken negative the 
system can no longer be given a simple interpretation as 
two-dimensional quantum gravity. However,
it can be shown that they can be given the interpretation as 
certain matter fields coupled to two-dimensional random lattices
of the kind used above in constructing the theory of 
two-dimensional quantum gravity.
The fine-tuning of the coupling constants give us new continuum 
systems which can be viewed as certain conformal field theories 
coupled to two-dimensional quantum gravity \cite{book,kazakov,staudacher}. 

Thus we can view these 
generalized systems as a school book example of critical systems:
they are defined in an infinite dimensional coupling constant space 
$g_3,g_4,\ldots$. Fine-tuning the coupling constants bring us 
to a critical hyper-surface of co-dimension one, describing 
two-dimensional quantum gravity, and further fine-tuning leads to 
the critical behavior of certain conformal field theories 
coupled to two-dimensional quantum gravity. The critical 
exponents of the conformal field theories are changed (relative to 
their value in flat two-dimensional space) due to the 
influence of the fluctuating geometry, and likewise the behavior 
of two-dimensional quantum gravity is changed by the back-reaction 
of matter (see \cite{book} for details).

\subsection{Quantum geodesic distance}

The critical behavior of the conformal field theories 
coupled to quantum gravity can be calculated by looking at 
globally defined matter correlators. Let $\phi(x)$ be a scalar matter field.
Then 
\beq\label{3.10}
\int \cD [g(x)] \cD \phi(x) \; \e^{-S[g,\phi]} \int \d^2 x\sqrt{g(x)} 
\int \d^2y \sqrt{g(y)} 
\; \phi(x) \phi(y)
\eeq
is reparameterization invariant and the scaling behavior of 
this correlator with respect to the cosmological constant $\L$ determines
the scaling dimension of the fields and the critical exponents. The
reason for this is that the average volume of space-time is
monitored by the cosmological constant,
\beq\label{3.11}
\la V \ra \sim \L^{-1},
\eeq
and  thus the finite-size scaling relations of the system are 
determined as functions of the cosmological constant.

However, from the theory of critical phenomena we know that the 
underlying universality of large scale fluctuations originating 
from quite different microscopic interactions is due to 
a {\it divergent} correlation length $\xi(g_3,g_4,\ldots)$.
How do we define such a length in quantum gravity? 
An object like $\la \phi(x) \phi(y)\ra$ has clearly no reparameterization
invariant meaning if we view $x$ and $y$ as coordinates. Further,
we have to integrate over all geometries and the 
(geodesic) distance between $x$ and $y$ will vary. 
One can define the concept of an invariance correlation function
as follows \cite{aw, ajw}: 
\bea\label{3.12}
\la \phi(D)\phi(0)\ra &\equiv&
\int \cD [g(z)] \cD \phi(z) \; \e^{-S[g,\phi]} \\
&& \int \d^2 x\sqrt{g(x)}\int\d^2 y \sqrt{g(y)}\; 
\del\Big(D_g(x,y)-D\Big )\, \phi(x) \phi(y). \nonumber
\eea 
In \rf{3.12} $D_g(x,y)$ denotes the {\it geodesic distance} between 
the space-time points labeled $x$ and $y$. 
Let us consider the simplest case where there are 
no matter Lagrangian at all and $\phi(x)\equ 1$ in \rf{3.12}.
In this case \rf{3.12} can be viewed as the partition function 
$G(D)$ for universes where two marked points are separated a given 
geodesic distance $D$. Again the calculation of this partition 
function reduces to the combinatorial problem of  counting 
a certain class of two-geometries. And again one can solve 
the combinatorial problem. 
In the regularized theory one obtains for $k_2$ close 
to the critical point \cite{aw}:
\beq\label{3.13}
G_a(d;k_2) = (k_2-k_2^c)^{3/4} \frac{\cosh d \sqrt[4]{k_2-k_2^c}\, d}
{\sinh^3 d \sqrt[4]{k_2-k_2^c}}.
\eeq
Here $d$ denotes a suitable discretized geodesic distance~\footnote{
One expects ``universality'' in such choice. For instance one can 
define a distance between two vertices in the triangulation as the 
smallest number of links connecting them. Other definitions lead 
to identical results (see \cite{ajw} for  details).}

If we take the continuum limit for the above ``two-point'' function
it is seen that we are forced to scale geodesic distance {\it anomalously}:
\beq\label{3.14}
D = d \sqrt{a}.
\eeq
With this scaling we obtain
\beq\label{3.15}
G(D;\L) = \L^{3/4} \frac{\cosh \sqrt[4]{\L} D}{\sinh^3 \sqrt[4]{\L} D}.
\eeq
This function behaves as two-point function: it falls of exponentially 
for large distances and it behaves power-like for distances
smaller that $1/\sqrt[4]{\L}$. 
The anomalous dimension of the geodesic distance 
is also reflected in the average space-time volume $V$ enclosed in disc of 
(geodesic) radius $D$. One finds
\beq\label{3.16}
\la V\ra_D \sim R^4,~~~~~~R \ll 1/\sqrt[4]{\L}.
\eeq
Thus the Hausdorff dimension of a typical two-dimensional geometry 
is four, as first realized in \cite{kkmw}. This is a genuine quantum 
phenomenon. 

Until now it has not been possible to calculate analytically the 
two-point function \rf{3.12} when matter fields are present. However,
it can be studied numerically and it can be verified that for 
the Ising model and the three-state Potts model coupled to two-dimensional
quantum gravity one really obtains a {\it divergent} spin-spin 
correlation length when the spin couplings approach their critical 
values (see \cite{ak,ak1} for details).

 \subsection{The proper-time propagator and quantum geometry}

\begin{figure}[t]
\epsfxsize=10pc 
\centerline{\epsfbox{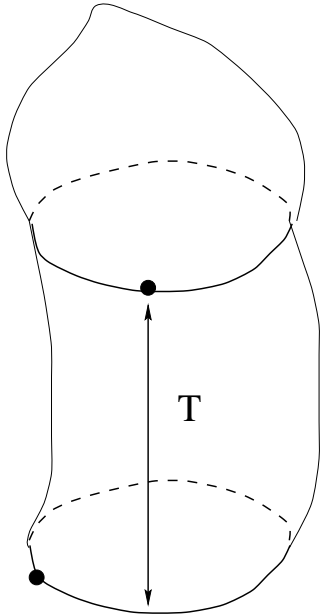}} 
\caption{A geometry contributing to the punctured disc amplitude (the 
punctured Hartle-Hawking
wave function), decomposed into geometries contributing to the 
proper-time propagator and the disc amplitude.  \label{figw}}
\end{figure}

\begin{figure}[t]
\epsfxsize=25pc 
\centerline{\epsfbox{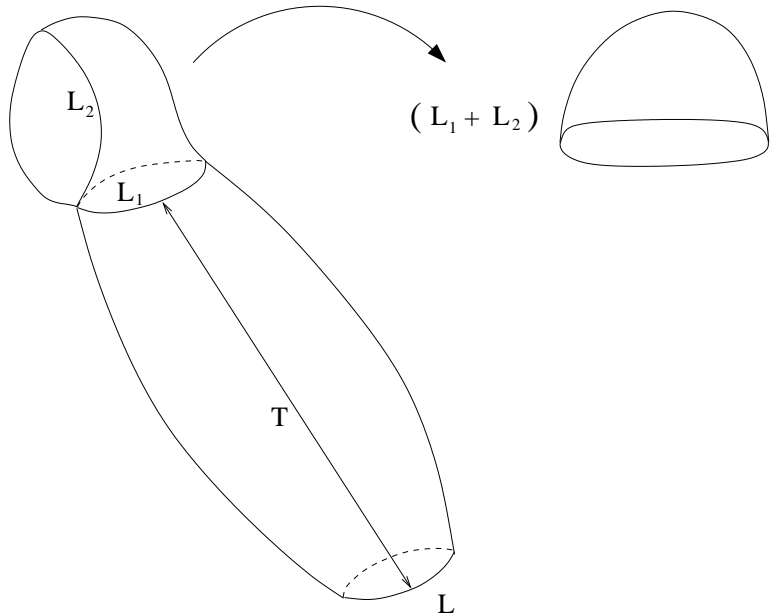}} 
\caption{A geometry contributing to $G(L,L_2;\L)$ decomposed 
into geometries contributing to $G_\L(L,L_1;T)$ and 
$G_{fig\mi 8}(L_1+L_2;\L)$. The latter is related to the disc amplitude
(the Hartle-Hawking wave function) $G(L_1+L_2;\L)$ as shown. \label{fig8}}
\end{figure}

An important tool used in the calculation of $G(D;\L)$ is the 
proper-time propagator. In Euclidean geometry proper-time $T$ is 
equivalent to geodesic distance, which we have denoted $D$. 
Let $B_1$ be a boundary. 
The distance of a point $P$ to $B_1$ is defined as the 
minimal geodesic distance of $P$ to the points of $B_1$.
We define another boundary $B_2$ to be separated from 
it by a geodesic distance 
$D$, or proper-time $T$, if  the distance of {\it each} point on $B_2$
to $B_1$ is $D$ (note that the definition is asymmetric with 
respect to $B_1$ and $B_2$). The proper-time propagator $G_\L(B_1,B_2;T)$
is defined by summing over all two-geometries with boundaries $B_1$ and 
$B_2$, such that $B_2$ has geodesic distance $T$ to $B_1$, the weight
of each configuration given as usually by $\e^{-S}$. 
It was first calculated in \cite{kkmw}, and in \cite{al1} it was 
shown that the equation which determines it 
can be obtained by very simple ``quantum geometric''
reasoning. Consider Fig.\ \ref{figw}. It shows a disc amplitude with 
one marked interior point $P$ (the marked point on the boundary is 
irrelevant for the considerations to follow) 
Since $P$ can be anywhere, marking a point 
corresponds to multiplying with the  space-time volume $V$ or 
(equivalently) differentiating after the cosmological constant.
However, one can also make the following decomposition: $P$
has a geodesic distance $T$ to the boundary ``1''. Form the connected loop
through $P$ which has distance $T$ to the boundary. One obtains all 
geometries with one marked point by integrating over $T$ and the 
length of the loop through $P$. We thus have the 
following consistency equation involving the Hartle-Hawking wave function and 
the proper-time propagator:
\beq\label{3.17}
\frac{\prt G(L_1;\L)}{\prt \L} = 
\int \d T \d L_2 \; G_\L (L_1,L_2;T) G(L_2;\L)
\eeq
Under reasonable scaling assumptions one can show that 
eq.\ \rf{3.17} determines the behavior of both 
$G(L;\L)$ and $G_\L(L_1,L_2;T)$. Quite surprisingly one finds {\it two} 
different solutions \cite{al1}. One scaling behavior leads to the 
Hartle-Hawking wave function already encountered. The other
consistent scaling behavior leads to a different two-dimensional
theory of quantum gravity which we call ``Lorentzian'' quantum gravity
for reasons which will become clear in Sec. \ref{lor}. At the moment we 
concentrate on the ``Euclidean'' quantum gravity theory discussed 
until now. Rather than discussing the explicit form of the 
proper-time propagator, let us give a further example of simple 
relations derived using ``quantum geometry'' (for more examples and 
a description of ``operator product expansions'' in two-dimensional 
quantum gravity see \cite{nishimura}). The example will 
illustrate that the proper-time propagator contains {\it all} 
information about quantum gravity.
We have already mentioned that the usual Hartle-Hawking wave function 
$G(L;\L)$ can be derived from properties of 
$G_\L (L_1,L_2;T)$. Let us consider 
instead the two-loop function $G(L_1,L_2;\L)$. It is not simply obtained 
from by integrating  $G_\L (L_1,L_2;T)$ over $T$ since 
boundary ``2'' of the space-times included in $G_\L (L_1,L_2;T)$
has a well-defined distance to boundary ``1''. No such relation exists 
between the boundaries in $G(L_1,L_2;\L)$. However, starting at boundary 
``1'' and moving in successive proper-time steps one will end in the 
situation shown in Fig.\ \ref{fig8}. 
The ``figure 8'' amplitude \cite{israel} 
with a boundary of length $L$ is related to the disc amplitude by
\beq\label{3.18}
G_{fig-8} (L;\L) \sim L G(L;\L),
\eeq
simply because it is obtained by pinching the boundary of the 
disc amplitude. Finally we can write:
\beq\label{3.19}
G(L_1,L_2;\L) = \int \d T \d L_3 \; G_\L (L_1,L_3;T)\, (L_2+L_3)G(L_2+L_3;\L).
\eeq

Like the expression for the disc amplitude, which could be derived 
from general properties of the proper-time propagator, \rf{3.19} is 
also an example of ``quantum geometric'' considerations, which in 
essence are just combinatorial identities. 

\subsection{Summary}

Euclidean two-dimensional quantum gravity as described above 
as the continuum limit of a regularized theory of 
two-dimensional geometries is still awaiting a precise 
mathematical description. What is meant by this is the 
following: Consider the random walk process in $R^d$. One can take 
the continuum limit in much the same way as we did in the 
construction of $G(L_1,\ldots,L_n;\L)$. The number of random walks
between  point $x$ and $y$ in $R^d$ grows exponentially with the 
number of steps in the random walk process and by fine-tuning the 
fugacity or stopping probability of the random walk process
one can obtain the relativistic scalar particle propagator
from point $x$ to point $y$. This is in fact the typical 
``first quantized'' path integral derivation of the free propagator.
In this case we know the correct mathematical measure on the 
space of continuous path from $x$ to $y$: it is the Wiener measure~\footnote{A
subtlety should be mentioned here: The usual Wiener measure is defined 
on the class of {\it parameterized} path. We need it on the class of 
{\it unparameterized} path. See \cite{book} for a detailed discussion.}.
We would like to construct the corresponding measure on the 
space of two-dimensional continuous geometries. It should exist!
Let us list the analogies between the the random walk and the ``random
geometries'': Their numbers grow exponentially with the number of 
``elements'' (steps in the random walk, triangles etc in the case of 
random geometries). The constants in the exponents are non-universal,
but the leading corrections to the exponential growth {\it are}
universal. A ``typical'' path or random geometry is fractal, in 
fact in almost the same way. The length of a ``typical'' path 
between $x$ and $y$ grows like $|x\mi y|^2$ rather than like $|x\mi y|$.
In the same way the area of a random two-geometry of 
(geodesic) diameter  $D$ grows as $D^4$ rather than $D^2$. 
It thus seems that the class of continuous two-dimensional geometries 
constitutes a well-defined class of geometries, not much ``wilder''
than the class of continuous path~\footnote{A more precise 
description of the geometries will be given below when Lorentzian 
quantum gravity is discussed}. Our results suggest that it 
should be possible to define a measure on this set, since we can 
constructively define the path integral over the set of 
two-geometries.

Some of the results mentioned above can be ``guessed'' from 
the continuum formulation of two-dimensional quantum gravity
known as ``Liouville quantum gravity''.  
Let us consider a fixed topology, for instance that of the 
sphere, which we represent as the complex plane with $\infty$ identified 
as the north-pole. Any metric can be 
decomposed as 
\beq\label{3.20}
g_{\m\n} = \e^{\phi}\del_{\m\n},
\eeq
and has to satisfy certain regularity conditions at $\infty$.
The corresponding classical action is simply 
\beq\label{3.21}
S = \L \int \d^2 x \sqrt{g} = \L \int \d^2 x\; \e^{\phi}.
\eeq
However, $\phi(x)$ is not a genuine dynamical field and the kinetic 
term comes entirely from the gauge fixing in the path integral:
\beq\label{3.21a}
Z(\L) = \int \cD \phi  \;\e^{-\int \d^2 x
(\L \e^{\phi} + \a (\prt \phi)^2 )}.
\eeq
There are two problems associated with this partition 
function. The first is that a cut-off $a$ should be 
invariant under reparameterization, i.e.\
\beq\label{3,22}
a^2 = \e^{\phi(x)} (dx_1^2+dx_2^2),
\eeq
and the cut-off in parameter space $R^2$ is thus field dependent. 
The next problem is 
the geometric observables we have discussed so far 
become non-local and complicated when formulated 
in terms of $\phi(x)$. For instant the concept of 
geodesic distance as used above in the combinatorial 
approach is difficult to treat in terms of 
$\phi(x)$. Nevertheless, a number of the results
from the combinatorial approach has been verified in 
quantum Liouville theory via the study of vertex operators
of the theory in the following sense: the study of 
the scaling properties of vertex operators allows one to make
reasonable guesses of the scaling behavior of the ``observables''
$G(L_1,\ldots,L_n;\L)$ and in this way recover the results 
obtained constructively by combinatorial methods 
(see for instance \cite{mss}). It is worth
emphasizing this, because it shows that there {\it is} an
underlying continuum quantum theory at the critical point
of the discretized theory, described by a Lagrangian and 
some standard rules of quantization, which however are technically 
difficult to implement because of the reparameterization invariance 
of the theory.

\section{Higher dimensions}\label{higherd}

Encouraged by the success of two-dimensional Euclidean quantum gravity
one can try to follow the program outlined in the 
introduction and look for a non-trivial fixed point 
of four-dimensional quantum gravity in the spirit of 
{\it asymptotic safety} as defined by Weinberg.
It {\it is} indeed possible to find  a candidate 
fixed point using the simplest action mentioned in Sec.\ \ref{nonpert}
 \cite{aj,am}.
This fixed point seemed originally to relate
in very interesting ways to various other proposals 
for a theory  quantum gravity \cite{amm,kkk}, but extensive computer
simulations eventually established that the 
phase transition was a (weakly) {\it first order} transition 
\cite{bielefeld} and 
thus could not serve as the desired fixed point of second 
or higher order~\footnote{The situation is slightly more complicated:
There is still an infinite volume limit for {\it all} values of the 
bare gravitational coupling constant by an appropriate fine-tuning 
of the cosmological coupling constant, as described in Sec.\ \ref{nonpert}. 
However, unlike in two dimensions
this infinite volume limit seems to be uninteresting from a 
physical point of view. We expect in four-dimensional quantum 
gravity to have genuine dynamical propagating degrees of freedom 
and the situation is thus more like two-dimensional quantum 
gravity coupled to for instance an Ising model. While we can always
take the infinite volume limit, there will only be one value of 
$\b$ (of $k_2$) where the Ising spin correlation length diverges
(where the graviton is massless).}.
In terms of the properties of ``typical'' configurations of
four-dimensional Euclidean geometries there is a sharp contrast 
with the two-dimensional case: a typical geometry, constructed 
by gluing together equilateral four-simplices with weight ``one''
is singular: the four-dimensional space-time volume grows at least 
exponentially 
with geodesic radius of the universe: Thus one cannot talk about 
a definite fractal dimension of a ``typical'' geometry. The 
first order phase transition observed in the numerical simulations
is a transition from such ``crumpled'' geometries of no extension
to another phase where the geometries are ``maximally elongated'' and 
fractal, of fractal dimension two, namely so-called branched polymers
\cite{aj1}.
Neither set of geometries can serve as an underlying 
set of geometries defining an interesting 
theory of four-dimensional quantum gravity. 
  
The original hope was that {\it at} the transition point one 
could define a set of geometries of a well defined fractal dimension 
larger than or equal to four, which could serve as the underlying set 
of geometries used in the definition of the continuum path integral.
However, the first order transition rules out this possibility: there is 
no smooth interpolation between to two extreme classes of geometries 
mentioned above.

No simple modification of the 
Einstein-Hilbert action seems to
change the order of the transition \cite{ajk}. The conclusion is that   
a continuum theory of Euclidean four-dimensional  
quantum gravity does not exist 
as a limit of the simplicial  formulation given here.

\section{Lorentzian quantum gravity}\label{lor}

\begin{figure}[t]
\epsfxsize=20pc 
\centerline{\epsfbox{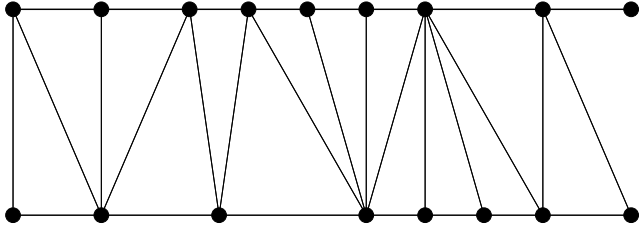}} 
\caption{Two successive spatial slices connected by fundamental 
building blocks. The spatial link of a  triangle 
has $ds^2= a^2$ while the (two) time-like links 
have $ds^2= -a^2$ before rotating to Euclidean space-time. \label{figlor}}
\end{figure}

The relation between the theory of critical phenomena and 
Euclidean quantum field theory is well established. 
However, if the field theory in question is ``quantum gravity''
the situation is unclear. The results for Euclidean gravity 
mentioned in the former Section indicate that  in higher than two 
dimensions one should not sum over Euclidean geometries if 
one wants a viable theory of quantum gravity. Twenty years ago 
Teitelboim suggested \cite{teitelboim} 
that one should only sum over causal space-time 
histories in the (Minkowskian) path integral if one wanted to 
maintain causality in the quantum theory of 
gravity~\footnote{Related ideas have been discussed in 
\cite{lee,sorkin}.}.
This idea was made concrete in simplicial quantum gravity 
in \cite{al1,ajl} and 
was denoted {\it Lorentzian simplicial quantum gravity}. More precisely,
consider the proper-time propagator already defined above
in the context of Euclidean quantum gravity. Let the allowed geometries
entering in the sum between two spatial boundaries, separated
by proper time $T$ be all non-degenerate causal geometries which allow a 
proper time slicing~\footnote{It is often stated that proper-time is 
a bad ``time'' choice since it can become singular in the 
context of initial value problems of General Relativity. However, in the 
path integral such singular behavior is most likely  unimportant and of zero 
measure.}. In the context of simplicial gravity this can be made 
precise (see \cite{al1} for details): starting with a given spatial 
geometry one defines constructively the class of space-time geometries 
with a boundary separated from initial spatial surface by proper-time 
$a$, $a$ been the lattice spacing, i.e.\ the cut-off. 
This is illustrated in the simplest case of two-dimensional Lorentzian 
gravity in Fig.\ \ref{figlor}, but can be generalized to three-- and 
four-dimensional cases \cite{ajl}.
Proceeding this 
way one defines the simplicial space-time geometries with two 
spatial boundaries separated a proper-time $T \equ a\cdot n$ and 
which allow a proper-time slicing. 

The class of simplicial geometries defined this way has the virtue
that each geometry allows a natural rotation to Euclidean signature.
Also the action is rotated as follows: $i S_M \to -S_E$, as in 
ordinary quantum field theory. This opens for the possibility to 
follow the standard procedure and first rotate to Euclidean signature,
then perform the summation over space-time histories and finally 
to rotate proper-time back to Minkowskian signature. The difference 
from the Euclidean path integral considered in Sec.\ \ref{higherd} 
is the requirement 
that each Euclidean geometry entering in the sum over space-time 
histories comes from a Minkowskian 
non-degenerate geometry which allows a proper-time slicing. In particular,
this implies that the spatial geometry obtained at a given proper-time
is always connected. This is quite natural from the 
point of view of  canonical quantization. The possibility of 
splitting space in several components is not natural in the 
framework of canonical quantization. However, if we return to 
two-dimensional Euclidean quantum gravity it is seen that 
the proper-time slicing does not respect this. Starting
out with a connected boundary of a ``typical'' two-dimensional 
Euclidean geometry and implementing  a proper-time slicing,
the spatial slices will in general consist of many disconnected 
components. In fact, the main difference between the class of 
of geometries used in Euclidean two-dimensional quantum gravity 
and Lorentzian two-dimensional quantum gravity can be understood 
by looking at the proper-time slicing \cite{ackl}. In Lorentzian 
quantum gravity the typical fluctuating geometry is genuining 
{\it two-dimensional}~\footnote{The fact that a typical Lorentzian geometry 
is not {\it fractal} does not mean that it is not fluctuating. In 
fact, in the case of two-dimensional Lorentzian geometries, the 
fluctuation of the spatial volume (length) $L(t)$, where $t$ denotes 
proper time, is maximal: $\la L^2\ra \mi \la L\ra^2 \sim \la L\ra^2$.}. 
One obtains a typical Euclidean geometry
by allowing the outgrowth of a baby universe at each space-time point.
In this way the four-dimensional fractal structure of a typical 
quantum configuration can be understood as a kind of  ``product'' of 
ordinary two-dimensional space-times.

\begin{figure}[t]
\epsfxsize=18pc 
\centerline{\epsfbox{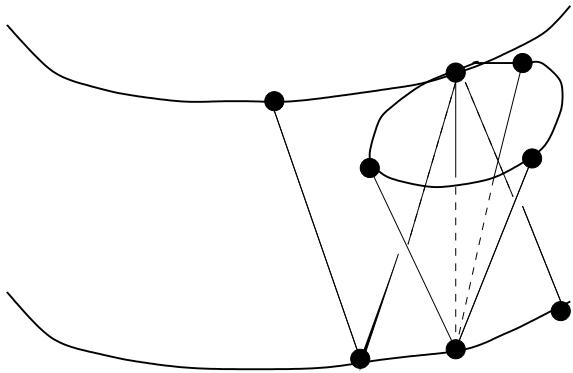}} 
\caption{The creation of a baby universe in the time-evolution
of two-dimensional space-times indicated in 
Fig.\ \ref{figlor}. At successive time-steps the baby universe 
should be considered as detached from the ``parent universe''. \label{figbaby}}
\end{figure}

In view of the universality mentioned in connection with 
two-dimensional Euclidean quantum gravity it is surprising
that there exists another fixed point belonging to a different 
universality class and corresponding to Lorentzian two-dimensional 
quantum gravity, but this universality class is now well understood
in the context of critical phenomena \cite{paris}. To make this 
possibility of two different types of critical behavior more 
clear let us consider a model which allows for the creation of 
baby universes along with the propagation of the two-dimensional 
Lorentzian space-time. Such a possibility is shown in 
Fig.\ \ref{figbaby}. Compared to Fig.\ \ref{figlor} one allows 
for the possibility of a baby universe to be created at any point at the 
spatial slice at proper time $t$ and then to develop independently 
of the ``parent universe''.
Let $G_\L(L_1,L_2;T)$ be the proper-time propagator mentioned above,
$T$ being the proper-time and let $G_\L(X,Y;T)$ be the Laplace 
transform of $G_\L(L_1,L_2;\L)$ (and similarly $G(X;\L)$ the 
Laplace transform of the Hartle-Hawking wave function $G(L;\L)$).
Again it has a combinatorial interpretation and from this one 
can derive the following equation (see \cite{al1} for details):
\beq\label{5.1}
a^\ep \frac{\prt}{\prt T}G_\L (X,Y;T) = 
-a  \frac{\prt}{\prt X} \Big( [X^2\mi \L \pl 
\k a^{\eta-2}G(X;\L)] G_\L(X,Y;T)\Big)
\eeq
Here $\k$ is the coupling constant for creating a baby universe 
at a space-time point. The exponents $\ep$ and $\eta$ are related 
to the dimension of proper-time (or geodesic distance) and the 
Hartle-Hawking wave function: $T \sim a^\ep$ and $G(X,\L) \sim a^{-\eta}$,
where $\sim$ just indicates the dimension. One can show, using 
eq.\ \ref{3.17} (i.e.\ Fig.\ \ref{figw}) that there are only 
two consistent choices: (1) $\k = 0$, $\ep\equ 1$, and 
$\k > 0$, $\ep \equ 1/2$ and $\eta \equ 3/2$ (see again \cite{al1}
for details). The first case corresponds 
to two-dimensional Lorentzian quantum gravity 
where proper-time scales canonically in 
accordance with the fact that the fractal dimension of  a 
typical geometry is two.
The second case corresponds to two-dimensional Euclidean quantum 
gravity where the geodesic distance scales anomalously and the fractal 
dimension of a typical geometry is four.
  
The coupling of matter to two-dimensional 
Lorentzian gravity is also (partially) understood. The Ising model
maintains its flat space critical exponents \cite{aal1}, and the interaction 
between matter fields and quantum gravity is weak~\footnote{Again it should 
be emphasized that a ``weak'' interaction does not imply that 
there is no interaction, only that it does not change general properties 
of the class of geometries used to define the continuum 
path integral, i.e. the fractal dimension of these geometries.} as long 
as the central change $c$ of the matter field is less than one~\footnote{When
$c > 1$ computer simulations have shown a strong back-reaction 
of matter on geometry \cite{aal2}.}.
This is in contrast to Euclidean quantum gravity where the 
critical exponents of matter as well as the critical exponents of 
quantum gravity itself are changed due to the coupling. 

\subsection{Higher dimensional Lorentzian quantum gravity}

The theory of Lorentzian simplicial quantum gravity 
has been formulated in two, three and four dimensions.
It has been solved analytically in two dimensions, as 
described above and its relation to two-dimensional Euclidean 
quantum gravity is understood.  The main question is of course 
if it provides us with a viable four dimensional theory of 
quantum gravity in the spirit of asymptotic safety. So far only 
the three-dimensional theory has been investigated in some detail,
this made possible because it can be mapped on a matrix model which 
allows us to do perform some analytical calculations \cite{ajlv} and because 
it is possible to perform numerical simulations \cite{ajl-num}.
Three-dimensional quantum gravity is not without interest. Although
it contains no propagating gravitons, i.e.\ no field-theoretical
degrees of freedom, it is formally non-renormalizable when 
expanded around a fixed background geometry. The results obtained
so far are quite encouraging. They indicate that one can take 
a continuum limit which describe quantum fluctuations around 
a background geometry created by the presence of a positive 
cosmological constant. However, more work is needed to substantiate 
this interpretation. If this works out and one can 
establish a firm connection to canonically quantized three-dimensional 
quantum gravity, Lorentzian simplicial quantum gravity will be 
a most interesting candidate for a regularized, background independent  
theory of four-dimensional quantum gravity, formulated entirely in 
terms of fluctuating geometries.

\section*{Acknowledgments}
It is a pleasure to thank K. Anagnostopoulos, J. Jurkiewics, R. Loll and
Y. Makeenko collaborations associated with the work 
reported here.
This work was supported by ``MaPhySto'', 
the Center of Mathematical Physics 
and Stochastics, financed by the 
National Danish Research Foundation as well as by 
the EU network on ``Discrete Random Geometry'', grant HPRN-CT-1999-00161, 
and by the ESF network no.82 on ``Geometry and Disorder''.

\end{document}